\documentstyle[aaai]{article}

\newtheorem{definition}{Definition}

\newtheorem{theorem}{Theorem}

\def\proof{\noindent{\bf Proof.}\enspace}
\def\sqr#1#2{{\vcenter{\hrule height.#2pt
      \hbox{\vrule width.#2pt height#1pt \kern#1pt
         \vrule width.#2pt}
      \hrule height.#2pt}}}
\def\square{\mathchoice\sqr34\sqr34\sqr{2.1}3\sqr{1.4}3}
\def\proofend{\nobreak\qquad$\square$\medskip}

\long\def\ignore#1{}

\newcount\xtemp
\newcount\ytemp
\newcount\xxtemp
\newcount\yytemp
\newcount\digl
\newcount\ddigl
\newcount\mdigl
\newcount\hdigl
\newcount\mhdigl
\newcount\mddigl
\hdigl= \digl
\divide\hdigl by 2
\mhdigl= \digl
\divide\mhdigl by 2
\multiply\mhdigl by -1
\ddigl= \digl
\multiply\ddigl by 2
\mdigl=\digl
\multiply\mdigl by -1
\mddigl=\ddigl
\multiply\mddigl by -1

\newcount\boxsize
\newcount\halfboxsize
\newcount\alhalfboxsize
\halfboxsize= \boxsize
\divide\halfboxsize by 2
\alhalfboxsize= \boxsize
\divide\alhalfboxsize by 2
\advance\alhalfboxsize by -2

\newcount\boxsize
\newcount\halfboxsize
\newcount\alhalfboxsize
\halfboxsize= \boxsize
\divide\halfboxsize by 2
\alhalfboxsize= \boxsize
\divide\alhalfboxsize by 2
\advance\alhalfboxsize by -2

\def\PDmatrix{
\begin{center}
\begin{picture}(60,60)(30,0)
\put (-7,11){\bf A}
\put (69,60){\bf B}
\put (10,20){
\begin{tabular}{c|c|c|}
& $\cal D$ & $\cal C$ \\
\hline
\begin{picture}(10,\boxsize)
\put (0,-2.5){\makebox(10,\boxsize){$\cal D$}}
\end{picture} &
\begin{picture}(\boxsize,\boxsize)
\put (0,0){\makebox(\boxsize,\boxsize)[bl]{$P$}}
\put (\halfboxsize,\alhalfboxsize){\makebox(\halfboxsize,\alhalfboxsize)
[r]{$P$}}
\end{picture} &
\begin{picture}(\boxsize,\boxsize)
\put (0,0){\makebox(\boxsize,\boxsize)[bl]{$T$}}
\put (\halfboxsize,\alhalfboxsize){\makebox(\halfboxsize,\alhalfboxsize)
[r]{$S$}}
\end{picture} \\
\hline
\begin{picture}(10,\boxsize)
\put (0,-2.5){\makebox(10,\boxsize){$\cal C$}}
\end{picture} &
\begin{picture}(\boxsize,\boxsize)
\put (0,0){\makebox(\boxsize,\boxsize)[bl]{$S$}}
\put (\halfboxsize,\alhalfboxsize){\makebox(\halfboxsize,\alhalfboxsize)
[r]{$T$}}
\end{picture} &
\begin{picture}(\boxsize,\boxsize)
\put (0,0){\makebox(\boxsize,\boxsize)[bl]{$R$}}
\put (\halfboxsize,\alhalfboxsize){\makebox(\halfboxsize,\alhalfboxsize)
[r]{$R$}}
\end{picture} \\
\hline
\end{tabular}
}
\end{picture}
\end{center}
}

\def\DCWmatrix{
\begin{center}
\begin{picture}(75,85)(40,-5)
\put (-10,11){\bf A}
\put (79,80){\bf B}
\put (10,20){
\begin{tabular}{c|c|c|c|}
& $\cal D$ & $\cal C$ & $\cal W$ \\
\hline
\begin{picture}(10,\boxsize)
\put (0,-2.5){\makebox(10,\boxsize){$\cal D$}}
\end{picture} &
\begin{picture}(\boxsize,\boxsize)
\put (0,0){\makebox(\boxsize,\boxsize)[bl]{$P$}}
\put (\halfboxsize,\alhalfboxsize){\makebox(\halfboxsize,\alhalfboxsize)
[r]{$P$}}
\end{picture} &
\begin{picture}(\boxsize,\boxsize)
\put (0,0){\makebox(\boxsize,\boxsize)[bl]{$T$}}
\put (\halfboxsize,\alhalfboxsize){\makebox(\halfboxsize,\alhalfboxsize)
[r]{$S$}}
\end{picture} &
\begin{picture}(\boxsize,\boxsize)
\put (0,0){\makebox(\boxsize,\boxsize){$0$}}
\end{picture} \\
\hline
\begin{picture}(10,\boxsize)
\put (0,-2.5){\makebox(10,\boxsize){$\cal C$}}
\end{picture} &
\begin{picture}(\boxsize,\boxsize)
\put (0,0){\makebox(\boxsize,\boxsize)[bl]{$S$}}
\put (\halfboxsize,\alhalfboxsize){\makebox(\halfboxsize,\alhalfboxsize)
[r]{$T$}}
\end{picture} &
\begin{picture}(\boxsize,\boxsize)
\put (0,0){\makebox(\boxsize,\boxsize)[bl]{$R$}}
\put (\halfboxsize,\alhalfboxsize){\makebox(\halfboxsize,\alhalfboxsize)
[r]{$R$}}
\end{picture} &
\begin{picture}(\boxsize,\boxsize)
\put (0,0){\makebox(\boxsize,\boxsize){$0$}}
\end{picture} \\
\hline
\begin{picture}(10,\boxsize)
\put (0,-2.5){\makebox(10,\boxsize){$\cal W$}}
\end{picture} &
\begin{picture}(\boxsize,\boxsize)
\put (0,0){\makebox(\boxsize,\boxsize){$0$}}
\end{picture} &
\begin{picture}(\boxsize,\boxsize)
\put (0,0){\makebox(\boxsize,\boxsize){$0$}}
\end{picture} &
\begin{picture}(\boxsize,\boxsize)
\put (0,0){\makebox(\boxsize,\boxsize){$H$}}
\end{picture} \\
\hline
\end{tabular}
}
\end{picture}
\end{center}
}

\def\DCWOmatrix{
\begin{center}
\begin{picture}(95,95)(55,-20)
\put (-10,11){\bf A}
\put (89,95){\bf B}
\put (10,20){
\begin{tabular}{c|c|c|c|c|}
& $\cal D$ & $\cal C$ & $\cal O$ & $\cal W$ \\
\hline
\begin{picture}(10,\boxsize)
\put (0,-2.5){\makebox(10,\boxsize){$\cal D$}}
\end{picture} &
\begin{picture}(\boxsize,\boxsize)
\put (0,0){\makebox(\boxsize,\boxsize)[bl]{$P$}}
\put (\halfboxsize,\alhalfboxsize){\makebox(\halfboxsize,\alhalfboxsize)
[r]{$P$}}
\end{picture} &
\begin{picture}(\boxsize,\boxsize)
\put (0,0){\makebox(\boxsize,\boxsize)[bl]{$T$}}
\put (\halfboxsize,\alhalfboxsize){\makebox(\halfboxsize,\alhalfboxsize)
[r]{$S$}}
\end{picture} &
\begin{picture}(\boxsize,\boxsize)
\put (0,0){\makebox(\boxsize,\boxsize){$Q$}}
\end{picture} &
\begin{picture}(\boxsize,\boxsize)
\put (0,0){\makebox(\boxsize,\boxsize){$0$}}
\end{picture} \\
\hline
\begin{picture}(10,\boxsize)
\put (0,-2.5){\makebox(10,\boxsize){$\cal C$}}
\end{picture} &
\begin{picture}(\boxsize,\boxsize)
\put (0,0){\makebox(\boxsize,\boxsize)[bl]{$S$}}
\put (\halfboxsize,\alhalfboxsize){\makebox(\halfboxsize,\alhalfboxsize)
[r]{$T$}}
\end{picture} &
\begin{picture}(\boxsize,\boxsize)
\put (0,0){\makebox(\boxsize,\boxsize)[bl]{$R$}}
\put (\halfboxsize,\alhalfboxsize){\makebox(\halfboxsize,\alhalfboxsize)
[r]{$R$}}
\end{picture} &
\begin{picture}(\boxsize,\boxsize)
\put (0,0){\makebox(\boxsize,\boxsize){$Q$}}
\end{picture} &
\begin{picture}(\boxsize,\boxsize)
\put (0,0){\makebox(\boxsize,\boxsize){$0$}}
\end{picture} \\
\hline
\begin{picture}(10,\boxsize)
\put (0,-2.5){\makebox(10,\boxsize){$\cal O$}}
\end{picture} &
\begin{picture}(\boxsize,\boxsize)
\put (0,0){\makebox(\boxsize,\boxsize){$Q$}}
\end{picture} &
\begin{picture}(\boxsize,\boxsize)
\put (0,0){\makebox(\boxsize,\boxsize){$Q$}}
\end{picture} &
\begin{picture}(\boxsize,\boxsize)
\put (0,0){\makebox(\boxsize,\boxsize){$\hat{Q}$}} 
\end{picture} &
\begin{picture}(\boxsize,\boxsize)
\put (0,0){\makebox(\boxsize,\boxsize){$0$}}
\end{picture} \\
\hline
\begin{picture}(10,\boxsize)
\put (0,-2.5){\makebox(10,\boxsize){$\cal W$}}
\end{picture} &
\begin{picture}(\boxsize,\boxsize)
\put (0,0){\makebox(\boxsize,\boxsize){$0$}}
\end{picture} &
\begin{picture}(\boxsize,\boxsize)
\put (0,0){\makebox(\boxsize,\boxsize){$0$}}
\end{picture} &
\begin{picture}(\boxsize,\boxsize)
\put (0,0){\makebox(\boxsize,\boxsize){$0$}}
\end{picture} &
\begin{picture}(\boxsize,\boxsize)
\put (0,0){\makebox(\boxsize,\boxsize){$H$}}
\end{picture} \\
\hline
\end{tabular}
}
\end{picture}
\end{center}
}

\def\proofend{\nobreak\qquad\hbox{$\square$}\medskip}

\def\cutwspace{\setlength{\parskip}{0pt}\setlength{\itemsep}{1pt}}

\long\def\comment#1{}

\title{Time and the Prisoner's Dilemma}

\author{Yishay Mor \and Jeffrey S. Rosenschein \\
Institute of Computer Science\\
Hebrew University\\
Givat Ram, Jerusalem, Israel\\
ph: 011-972-2-658-5353\\
fax: 011-972-2-658-5439\\
email: yish@cs.huji.ac.il, jeff@cs.huji.ac.il }
        
\begin{document}

\maketitle

\begin{abstract}
This paper examines the integration of computational complexity into
game theoretic models. The example focused on is the Prisoner's
Dilemma, repeated for a finite length of time. We show that a minimal
bound on the players' computational ability is sufficient to enable
cooperative behavior.

In addition, a variant of the repeated Prisoner's Dilemma game is
suggested, in which players have the choice of opting out. This
modification enriches the game and suggests dominance of cooperative
strategies.

Competitive analysis is suggested as a tool for investigating
sub-optimal (but computationally tractable) strategies and game
theoretic models in general. Using competitive analysis, it is shown
that for bounded players, a sub-optimal strategy might be the optimal
choice, given resource limitations.

\end{abstract}




\section{Introduction}
\label{intro}

Alice and Bob have been arrested as suspects for murder, and are
interrogated in separate rooms. If they both admit to the crime, they
get 15 years of imprisonment. If both do not admit to the crime, they
can only be convicted for a lesser crime, and get 3 years
each. However, if one of them admits and the other does not, the
defector becomes a state's witness and is released, while the other
serves 20 years.

This is a {\bf ``Prisoner's Dilemma'' (PD)} game, a type of
interaction that has been widely studied in Political Science, the
Social Sciences, Philosophy, Biology, Computer Science, and of course
in Game Theory.  The feature of PD that makes it so interesting is
that it is analogous to many situations of interaction between
autonomous parties. The PD game models most situations in which both
parties can benefit from playing cooperatively, but each party can get
a higher gain from not cooperating when the opponent does. See
Figure~\ref{PDmat}.

\begin{figure}
\PDmatrix
\begin{center}
\(T > R > P > S\) \\
\(2R > T + S\)
\end{center}
\caption{\bf Prisoner's Dilemma game matrix}
\label{PDmat}
\end{figure}

As an example, consider two software agents, $A$ and $B$, sent by
their masters onto the Internet to find as many articles about PD as
they can. The agents meet, and identify that they have a common
goal. Each agent can benefit from receiving information from the
other, but sending information has a cost. The agents agree to send
packets of information to each other simultaneously. Assume that
sending an empty packet costs \$1, sending a useful packet costs \$2,
and receiving a useful packet is worth \$3.  This interaction is
precisely a PD game with $S = -2$, $P = -1$, $R = 1$ and $T =
2$.

Under the assumption of rationality, the PD game has only one
equilibrium: both players defect. This result is valid even for any
finite sequence of games---both players can deduce that the opponent
will defect in the last round, therefore they cannot be ``punished''
for defecting in the one-before-last round, and by backward induction
it becomes common knowledge that both players will defect in every
round of the repeated game.

The ``always defect'' equilibrium of PD is in a sense paradoxical; it
contradicts some of our basic intuitions about intelligent behavior,
and stands in contrast to psychological evidence~\cite{Rapoport62}.
The root of this paradox is the assumption of rationality, which
implies unlimited computational power; it is precisely the unlimited
computational power of rational agents that both allows and requires
them to perform the unlimited backward induction in the repeated
PD. In reality, both natural and artificial agents have limited
resources. In this paper we show that once these limitations are
incorporated into the interaction, cooperative behavior becomes
possible and reasonable.

The idea of bounding agents' rationality is not new. The novelty of
the approach presented here is in its straightforwardness. The bound
on rationality is measured in the most standard scale of Computer
Science: computation time. We assume that agents need time for
computations, and that the game is repeated for a finite length of
time, rather then a fixed number of iterations. These assumptions are
sufficient to create cooperative equilibria.

These results are interesting from two points of view, the system (or
environment) designer's perspective, and the agent designer's
perspective. From the system designer's point of view, it gives
guidelines as to how to create a cooperation-encouraging
environment. From the agent designer's point of view, it enables him
to design a strategy that will impose cooperation on his agent's opponent.

\subsection{Related Work}
\label{related}

A thorough and comprehensive survey of the basic literature on bounded
rationality and repeated PD appears in~\cite{Kalai90}.
Axelrod~\cite{Axelrod81,Axelrod84} reports on his famous computer
tournament and analyzes social systems accordingly.

Most of the work on this subject has centered on various automata as
models of bounded rationality;
\cite{Papadimitriou92,Gilboa89,Fortnow94} and others
(see~\cite{Kalai90} for an extensive bibliography) deal with finite
state automata with a limited number of states, and~\cite{Megiddo86}
examines Turing machines with a bounded number of states. The main
drawback of the automata approach is that cooperative behavior is
usually achieved by ``exhausting'' the machine---designing a game
pattern that is so complex that the machine has to use all its
computational power to follow it.  Such a pattern is highly non-robust
and will collapse in the presence of noise.

Papadimitriou, in~\cite{Papadimitriou92}, analyzes a 3-player variant
of the PD game. This game is played in two stages: first, every player
chooses a partner, then if two players choose each other, they play
PD. Two sections of this paper deal with a similar variation on the
repeated PD game, in which players have the possibility of opting out.


Several researchers (see~\cite{NowSig93,Binmore94} for
examples and further bibliography) took Axelrod's lead and
investigated evolutionary models of games. Most, if not all, of these
works studied populations of deterministic or stochastic automata with
a small number of states. The reason for limiting the class of
computational models investigated was mainly pragmatic: the
researchers had limited resources of computer space and time at their
disposal. We hope that this paper might give a sounder motivation for
the focus on ``simple'' or ``fast'' strategies. We claim that the same
limitations that hold for researchers hold for any decision maker, and
should be treated as an inherent aspect of the domain.

The task of finding papers on the Internet, as presented in the
example above, can be seen as a distributed search problem. The power
of cooperation in such problems has been studied in~\cite{Hogg93}.

Other examples of domains for which work of the type presented here is
relevant can be found in~\cite{Fikes94} and in~\cite{Foner94}. Foner's
system explicitly relies on autonomous agents' cooperation. He
describes a scenario in which agents post ads on a network,
advertising their wish to buy or sell some item. Ads (or, in fact, the
agents behind them) find partners by communicating with other agents,
and sharing with them information on ad-location acquired in previous
encounters. As Foner himself states:
\begin{quote} 
Such behavior is altruistic, in that any given agent has little incentive
to remember prior ads, though if there is uniformity in the programming
of each agent, the community as a whole will benefit, and so will each
individual agent.
\end{quote}
Game theory predicts that such behavior will not prevail. The results 
presented in this paper suggest that the computational incompetence of
the agents can be utilized to make sure that it does.

\subsection{Outline of the Paper}
\label{outline}

The section ``Finite Time Repeated Prisoner's Dilemma'' presents and
examines the finite time repeated PD game. The main result of this
section is in Theorem~\ref{CoOpEqStrong}, which shows weak conditions
for the existence of a cooperative equilibrium.

The following section introduces the possibility of {\em opting out},
parting from an unsuitable partner. In that section we mainly develop
tools for dealing with opting out, and show conditions under which
opting out is a rational choice.

In the section ``Sub-Optimal Strategies'' we use {\em competitive
analysis\/} to show that, in a sense, opting out strengthens the
cooperative players. Without it, a noncooperative player can force his
opponent into noncooperative behavior. The possibility of opting out
changes the balance of forces, allowing a cooperative player to force
an opponent into cooperative behavior.

\section{Finite Time Repeated Prisoner's Dilemma}
\label{time}

In this section we deal with a game of PD that is repeated for a
finite time (which we call the {\em FTPD\/} game). Previous
work~\cite{Kalai90} focused on finite or infinite iterated PD ({\em
IPD}). The basic idea is that two players play PD for N rounds.  In
each round, once both have made their move (effectively
simultaneously), they get the payoffs defined by the PD game
matrix. In our version of the game, the players play PD for a fixed
amount of (discrete) time. At each tick of the clock, if both made a
move, they get the PD payoff. However, if either player did not make
his move yet, nothing happens (both players get 0). See
Figure~\ref{FTPDmat}. 
Readers who are familiar with the game theoretic literature on PD,
might notice a problem that this perturbation entails.  If $H=0$, it
creates a new equilibrium point, thus technically eliminating the
paradox (this is, in fact, why the payoff when both players wait is
labeled $H$ and not defined as $0$). However, this is a technical
problem, and can be overcome by technical means, for instance by
setting $P=0$ or $H= -\epsilon$.  See~\cite{MyMast} for further
discussion.

\begin{figure}[htpb]
\DCWmatrix
\[ H \leq 0 \]
\caption[DCWmatrix]{\bf FTPD Game Payoff Matrix}
\label{FTPDmat}
\end{figure}

{\bf Rules of the FTPD Game:}
\begin{itemize}
\cutwspace
\item $2$ players play PD repeatedly for $N$ clock ticks, $N$ given as
input to players.
\item At each round, players can choose $C$ (cooperate), or $D$ (defect).
If they choose neither, $W$ (wait) is chosen for them by default.
\item The payoff for each player is his total payoff over $N$ rounds
(clock ticks).
\end{itemize}

\begin{theorem}
\label{unboundRatio}
If all players are (unboundedly) rational, and \(P >0 >H\), the FTPD game
is reduced to the standard IPD game.
\end{theorem}

\proof
The $W$ row (column) is dominated by the $D$ row (column), and thus 
can be eliminated.
\proofend

We now proceed to define our notion of bounded rationality, and
examine its influence on the game's outcome. Theorem~\ref{CBvsTuring}
is presented to enable the reader to compare our approach to the more
standard one of automata models.

\begin{definition}
\label{CBdef}
A {\bf Complexity Bounded (CB)} player is one with the following bound
on rationality: each ``compare'' action takes the player (at least)
one clock tick.\footnote{Formally, we have to say explicitly what we
mean by a ``compare''. The key idea is that any computational process
takes time for the player. Technically, we will say that a CB player
can perform at most $k$ binary $XOR$s in one clock tick, and $k <
\log_2{N}$.}
\end{definition}

Unless mentioned otherwise, we will assume this is the only bound on
players' rationality, and that each ``compare'' requires exactly one
clock tick.

\begin{theorem}
\label{CBvsTuring}
The complexity bound on rationality is weaker than restricting players
to Turing Machines. That is to say, any strategy realizable by a
Turing Machine can be played by CB players.
\end{theorem}

\proof Assuming that every read/write a Turing Machine
performs takes one clock tick, a Turing Machine is by definition
complexity bounded.
\proofend

{From} now on we will deal only with complexity bounded players.
Furthermore, we must limit either design time or memory to be finite
(or enumerable).\footnote{Papadimitriou~\cite{Papadimitriou92} makes a
distinction between design complexity and decision complexity. In our
model, decision complexity forces a player to play W, while design
complexity is not yet handled.  The choice of strategy is made at
design time; switching from $S_a$ to $S_b$ at decision time can be
phrased as a strategy $S_c$: ``play $S_a$, if\ldots switch to $S_b$''.}
Otherwise, our bound on rationality becomes void: a player can ``write
down'' strategies for any $N$ beforehand, and ``jump'' to the suitable
one as soon as it receives $N$.

The main objective of this section is to show how the complexity bound
on rationality leads to the possibility of cooperative behavior. To do
this, we must first define the concepts of equilibrium and
cooperativeness.

\begin{definition}
\label{Nash}
A {\bf Nash equilibrium} in an $n$-player game is a set of strategies,
$\Sigma = \{\sigma^1,..\sigma^n\}$, such that, given that for all $i$
player $i$ plays $\sigma^i$, no player $j$ can get a higher payoff by
playing a strategy other then $\sigma^j$.
\end{definition}

\begin{definition}
\label{CoOpEq}
A {\bf Cooperative Equilibrium} is a pair of strategies in Nash
equilibrium, such that, when played one against the other, they will
result in a payoff of \(R*N\) to both players.
\end{definition}

\begin{definition}
\label{CoOpStr}
A {\bf Cooperative Strategy} is one that participates in a cooperative
equilibrium.
\end{definition}

\begin{theorem}
\label{CoOpDefects}
A cooperative player will Wait or Defect only if his opponent Defected
or Waited at an earlier stage of the game.
\end{theorem}

\proof
If the player is playing against its counterpart in the cooperative
equilibrium and it Waits, its payoff is no more then $(N-1)*R$, in
contradiction to the definition of cooperative equilibrium.

For any other player, as long as the opponent plays $C$, the
cooperative player cannot distinguish it from its equilibrium
counterpart.
\proofend

Theorem~\ref{CoOpEqStrong} is the main result of this section.

\begin{theorem}
\label{CoOpEqStrong}
If \(R > 0\), then there exists a cooperative equilibrium of the FTPD
game.
\end{theorem}

\proof
Consider the strategy GRIM: 
\begin{enumerate}
\cutwspace
\item Begin by playing $C$, continue doing so as long as the opponent
does.
\item If the opponent plays anything other then $C$, switch to playing
$D$ for the remainder of the game.\footnote{\label{WatchOpp}The
strategy GRIM assumes a player can react to a Wait or Defect in the
next round, i.e., without waiting himself. This can be done, for
example, if the strategy is stated as an ``if'' statement: ``If opponent
played C, play C, else, play D.'' This strategy requires one ``compare''
per round.}
\end{enumerate}
Note that GRIM requires one ``compare'' per round, so having it played by 
both players results in N rounds played, and a payoff of \(N*R\) for each 
player.

Assume that both players decide to play GRIM. We have to show that no
player can gain by changing his strategy.

If player $A$ plays $D$ in round $k$, \(k < N\), then player $B$ plays
$D$ from that round on. $A$ can gain from playing $D$ if and only if
he plays $D$ in the $N$th round only. In order to do so, he has to be
able to count to $N$.  Since $N$ is given as input at the beginning of
the game, A can't design a strategy that plays GRIM $N-1$ rounds and
then plays $D$. He must compare some counter to $N-2$ before he
defects, to make sure he gains by defection. Doing so, he causes $B$
to switch to playing $D$. Therefore, his maximal payoff is:
\[ (N-1)*R + (P -R) \]
A will change his strategy if and only if \(P - 2R > 0\).
We assumed that \(R > P\).
If \(R > 0\), then \(2R > R\), and A will not switch.
Else, by assumption, \(2R > T + S\),
so A will switch only if \(P > T + S\),
and will not switch if \(T - P > - S\).
\proofend

\section{Finite Time Repeated Prisoner's Dilemma with Opting Out}
\label{optout}

In this section we study a variant of the FTPD game, which we call
OPD, in which players have the option of Opting Out---initiating a
change of opponent (see Figure~\ref{OPDmat}). A similar idea appears
in~\cite{Papadimitriou92}. While in the previous section we only
altered the nature of the players and the concept of iteration, here
we change the rules of the game. This requires some justification.

\begin{figure}
\DCWOmatrix
\caption[DCWOMatrix]{\bf OPD Game Matrix}
\label{OPDmat}
\end{figure}

The motivation for OPD is that it allows players greater flexibility
in choosing strategies. Consider player $A$ whose opponent plays
GRIM. In IPD or FTPD, once $A$ defects, the opponent will defect
forever after. From this point on, the only rational strategy for $A$
is also to defect until the end of the game. The possibility of opting
out enables $A$ to return to a cooperative equilibrium. Generally, the
existence of ``vengeful strategies'' (like GRIM) is problematic in the
standard game context. Other researchers~\cite{Gilboa89,Fortnow94}
have dealt with these strategies by explicitly removing them from the
set of strategies under consideration. Once opting out is introduced,
this is no longer necessary.

The possibility of opting out also makes the game less vulnerable to
noise, and provides fertile ground for studying learning in the PD
game context.  These issues are beyond the scope of the current paper;
see~\cite{MyMast} for further discussion.

One last motivation for this line of work is sociological: ``breaking
up a relationship'' is a common way of punishing defectors in repeated
human interactions. 
Vanberg and Congleton, in~\cite{Vanberg92}, claim that opting out
(``Exit'' in their terminology) is the {\em moral\/} choice, and
show that Opt-for-Tat is more successful than Tit-for-Tat in
Axelrod-type tournaments~\cite{Axelrod84}.

We start by comparing the OPD game to traditional approaches, namely
to games played by rational players and to the IPD game.

\begin{theorem}
\label{UnBoundOpt}
If all players are (unboundedly) rational, and \(P \geq Q \geq 0\)
(and $\hat{Q} < 0$, $H < 0$),
the OPD game is reduced to the standard IPD game.
\end{theorem}

\proof
The $W$ row (column) is dominated by the $D$ row (column), and thus 
can be eliminated.

Playing $O$ will result in a payoff of $Q$, \(Q < P\), and a switch in
partner. Since all players are rational, this is equivalent to
remaining with the same partner. A player cannot get a higher payoff
by playing $O$, hence the $O$ row and column are eliminated.
\proofend

\begin{theorem}
\label{FinRepOpt}
In the IPD game with opting out, if $P > Q > \hat{Q}$ and $H < 0$, then 
the only Nash equilibrium is \(<D^N,D^N>\).
\end{theorem}

\proof
The standard reasoning of backward induction works in this game;
see~\cite{MyMast} for the full proof.
\proofend 

{From} now on we will deal only with the OPD game. For simplicity's
sake, we will assume $Q = 0$, and $\hat{Q}=H=-\epsilon$.\\
Let us define the full context of the game.
\vskip 5pt
{\bf Rules of the OPD Game:}
\begin{enumerate}
\cutwspace
\item A population of $2K$ players is divided into pairs.
\item For every pair of players $<i,j>$, every clock tick, each player
   outputs an action $\alpha$, $\alpha \in \{C,D,O\}$. If he does not 
   output any of these, $W$ is assigned by default.
\item If either outputs $O$, the pair is split and both get $Q$, 
  regardless of the other player's action. Otherwise, if both play $C$ 
  or $D$, they get the PD payoff, and continue playing with one another.
  In any other case, both get 0 and remain paired.
\item Both players can observe their payoff for the previous round. We
   assume both do, and therefore ignore this monitoring action in our
   computations, i.e., assume this is done in 0 time.
\item Every $t$ clock ticks, all unpaired players are randomly matched.
\item The payoff to a player is the sum of payoffs he gets over $N$ clock
   ticks.
\end{enumerate}

We can now make a qualitative statement: opting out can be a rational
response to a defection. This is the intuition behind
Theorem~\ref{CoOpHigherPayoff}. In the next section we will
attempt to quantify this claim.

\begin{theorem}
\label{CoOpHigherPayoff}
The expected payoff when playing against a cooperative opponent is
higher than when playing against an unknown one.
\end{theorem}

\proof (Sketch - See~\cite{MyMast} for a more detailed proof.)      
Whatever $A$'s move is for the first round, his payoff against a
cooperative opponent is at least as high as against an unknown one.
If he plays anything other then $C$, his opponent becomes unknown, and
the proof is complete. If he plays $C$, we continue by
induction.
\proofend

\begin{theorem}
\label{InstRMtch}
If there is a positive probability of at least one of the other
players being cooperative, and rematching is instantaneous, then a
player in the OPD game has a positive expected gain from opting out
whenever the opponent waits.\footnote{The problematic condition is
instantaneous rematching. There are 2 necessary and sufficient
conditions for this to happen:
\begin{enumerate}
\cutwspace
\item Opting out can be done in the same round the opponent waits.
\item If a player opted out in this round, he will be playing against a
(possibly new) opponent in the next round, i.e., there is no ``transition
time'' from one opponent to another.
\end{enumerate}
The second condition is unjustifiable in any realistic setting.
The first condition returns to a player's ability to ``watch the
opponent'' without waiting, mentioned in Footnote~\ref{WatchOpp}.
In~\cite{MyMast} we show different assumptions that make this condition
possible.}
\end{theorem}

\proof
This theorem follows directly from Theorem~\ref{CoOpDefects} and
Theorem~\ref{CoOpHigherPayoff}. The full proof can be found
in~\cite{MyMast}.
\proofend

\section{Sub-Optimal Strategies}
\label{CompAnal}

In considering whether to opt out or not, player $A$ has to assess his
expected payoff against his current opponent $B$, the probability $B$
will opt out given $A$'s actions, and his expected payoff after opting
out. Such calculations require extensive computational resources, and
thus carry a high cost for a CB player. Furthermore, they require vast
amounts of prior knowledge about the different players in the
population. Although complete information is a standard assumption in
many game theoretic paradigms, it is infeasible in most realistic
applications to form a complete probabilistic description of the
domain.

As an alternative to the optimizing approach, we examine satisfying
strategies.\footnote{Herbert Simon~\cite{Simon69,Simon83} coined the
term ``Satisficing'' as an alternative to ``Maximizing.'' Although our
approach is close in spirit to his, it differs in its
formalism. Therefore we prefer to use a slightly different term.}
Instead of maximizing their expected payoff, satisfying players
maximize their worst case payoff. The intuition behind this is, that
if maximizing expected payoff is too expensive computationally, the
next best thing to do is to ensure the highest possible ``security
level,'' protecting oneself best against the worst (max-min).

In order to evaluate satisfying strategies, we use the method of
competitive analysis. Kaniel~\cite{Kaniel94} names Sleator and
Tarjan~\cite{Sleator84} as the initiators of this approach. The idea
is to use the ratio between the satisfying strategy's payoff and that
of a maximizing strategy as a quantifier of the satisfying strategy's
performance.
\vskip 5pt
We begin by defining the concepts introduced above.

\begin{definition}
\label{SecLev}
The {\bf Security Level} of a strategy $S$ is the lowest payoff a player
playing $S$ might get. Formally, if $\Gamma$ is the set of all possible
populations of players, and $\mu(S)$ is the expected payoff of $S$ then
the security level $SL(S)$ is:
\begin{center}
\large
min$_{\gamma\in\Gamma}$ \{$\mu |$ the population is $\gamma$\}
\end{center}
\end{definition}

\begin{definition}
\label{MaxNSat} \
\begin{itemize}
\cutwspace
\item A {\bf Maximizing} player is one that plays in a way that
maximizes his expected payoff.
\item A {\bf Satisfying} player is one that plays in such a way
that maximizes his security level.
\end{itemize}
\end{definition}

\begin{definition}
Let $A$ be a satisfying player, and let $S$ be $A$'s strategy. Let $h(S)$
be the expected payoff of $A$, had he been a maximizing player. The
{\bf Competitive ratio} of $S$ is:
\[       CR(S) =  {\frac{SL(S)}{h(S)}}.
\]
\end{definition}

The following two theorems attempt to justify examination of
satisfying rather than maximizing strategies.

\begin{theorem}
\label{OFTSatis}
If there is a probability of at least $q > 0$ of being matched with a
cooperative player at any stage of the game, then a player in the OPD
game can ensure himself a security level of $N*R -
const$.
\end{theorem}

\proof
Consider the strategy Opt-for-Tat (OFT). This is the strategy of
opting out whenever the opponent does not cooperate and cooperating
otherwise.  The expected number of times a player $A$ playing OFT will
opt out is $\frac{1}{q}$, after which he will be matched with a
cooperative player, and receive $R$ for each remaining round of the
game.

Actually, it is easy to show that $const =\frac{1}{q}*[(r +1)R -S]$ 
where $r$ is the expected number of rounds a player has to wait for a 
rematch~\cite{MyMast}.
\proofend

\begin{theorem}
\label{MaxPay}
If there is a probability of at least $q > 0$ of being matched with a
cooperative player, and all players in the population are satisfying or
optimizing players, a player in the OPD game cannot receive a payoff 
higher then $N*R + const.$
\end{theorem}

\proof
Assume there exists a strategy $\theta$ that offers a player $A$ playing 
it a payoff greater then $N*(R + \beta)$, $\beta > 0$. This means that in
some rounds $A$ defects and receives $T$. However, as soon as $A$ defects, 
he identifies himself as a $\theta$ player. His opponent can infer that 
playing against $A$ he will receive a payoff lower then the security level 
$N*R - const$, and will opt out.\\
Let $r$ be the expected number of rounds a player waits for a rematch.
If  $(r + 1)*R < T$ then $A$ loses every time he defects, and will get
a payoff $< N*R$. If $(r + 1)*R > T$ then ``defect always'' is the only
equilibrium strategy, and the probability of being rematched with a 
cooperative player becomes 0 (in contradiction to our assumption).
\proofend

{From} Theorems~\ref{OFTSatis} and~\ref{MaxPay} we get that as $N$ and
$q$ grow, the competitive ratio of a satisfying strategy in this game
approaches 1. If the time needed to compute the optimizing strategy is
proportional to $N$, a satisfying strategy is {\em de facto\/} optimal
for a CB player.

\subsection{Related work revisited}
The possibility of opting out makes cooperative, non-vengeful
strategies even stronger. This tool can be further developed into a
strategy-designing tool~\cite{MyMast}. We wish to demonstrate this 
claim using the examples presented in the introduction of this paper.

In domains like the paper-searching agents or Foner's ad-agents, 
cooperative information sharing is a desirable, if not required,
behavior. We propose the following guidelines for designers of such 
an environment:
\begin{itemize}
\item Ensure a large enough initial proportion of cooperative agents
in the domain (e.g., by placing them there as part of the system). Make
the existence of these agents known to the users of the system.
\item Advise the users to use the following protocol in informational
transactions:
\begin{itemize}
\item Split the information being transferred to small packets.
\item Use an OFT strategy, i.e., keep on sending packets as long as the
opponent does, break up and search for a new opponent as soon as a packet
does not arrive in time.
\end{itemize}
\item Inform the users of the satisfying properties of this strategy.
\end{itemize}
As shown in the section ``Sub-Optimal Strategies,'' it is reasonable
to assume that a large proportion of the users will implement
cooperative strategies in their agents.

\section{Conclusions}
\label{conc}

We introduced the {\em finite time repeated\/} PD game, and the notion
of {\em complexity bounded\/} players. In doing so, we encapsulated
both the player's utility and his inductive power into one parameter:
his payoff in the game. Cooperative equilibria arise as an almost
inherent characteristic of this model, as can be seen in
Theorem~\ref{CoOpEqStrong}. Furthermore, the common knowledge of
limited computational power enables an agent to control his opponent's
strategy. If the opponent spends too much time on computations, he is
probably planning to defect.

In the sections that followed, we introduced and studied {\em opting
out\/} in the PD game. We discussed both theoretical and intuitive
motivations for this variation on the standard game description. In
the section ``Sub-Optimal Strategies,'' we used the tool of {\em
competitive analysis\/} to show that the possibility of opting out
makes cooperative, non-vengeful strategies even stronger. We then
demonstrated the usefulness of these results with relation to the
examples presented in the introduction.

\nocite{KMRW}
\nocite{Axelrod84}

\section*{Acknowledgments}

This research has been partially supported by the Israeli Ministry of
Science and Technology (Grant 032-8284) and by the Israel Science
Foundation (Grant 032-7517).

\bibliographystyle{aaai}
\bibliography{icmas95}

\begin{thebibliography}{}

\bibitem[\protect\citeauthoryear{Axelrod \& Hamilton}{1981}]{Axelrod81}
Axelrod, R., and Hamilton, W.
\newblock 1981.
\newblock The evolution of cooperation.
\newblock {\em Science} 211(4489):1390--1396.

\bibitem[\protect\citeauthoryear{Axelrod}{1984}]{Axelrod84}
Axelrod, R.
\newblock 1984.
\newblock {\em The Evolution of Cooperation}.
\newblock New York: Basic Books.

\bibitem[\protect\citeauthoryear{Binmore \& Samuelson}{1994}]{Binmore94}
Binmore, K., and Samuelson, L.
\newblock 1994.
\newblock Drifting to equilibrium.
\newblock Unpublished manuscript.

\bibitem[\protect\citeauthoryear{Fikes \bgroup \em et al.\egroup
  }{1995}]{Fikes94}
Fikes, R.; Engelmore, R.; Farquhar, A.; and Pratt, W.
\newblock 1995.
\newblock Network-based information brokers.
\newblock In {\em The AAAI Spring Workshop on Information Gathering from
  Distributed, Heterogeneous Environments}.

\bibitem[\protect\citeauthoryear{Foner}{1995}]{Foner94}
Foner, L.~N.
\newblock 1995.
\newblock Clustering and information sharing in an ecology of cooperating
  agents.
\newblock In {\em The AAAI Spring Workshop on Information Gathering from
  Distributed, Heterogeneous Environments}.

\bibitem[\protect\citeauthoryear{Fortnow \& Whang}{1994}]{Fortnow94}
Fortnow, L., and Whang, D.
\newblock 1994.
\newblock Optimality and domination in repeated games with bounded players.
\newblock Technical report, Department of Computer Science University of
  Chicago, Chicago.

\bibitem[\protect\citeauthoryear{Gilboa \& Samet}{1989}]{Gilboa89}
Gilboa, I., and Samet, D.
\newblock 1989.
\newblock Bounded vs. unbounded rationality: The tyranny of the weak.
\newblock {\em Games and Economic Behavior} 1:213--221.

\bibitem[\protect\citeauthoryear{Hogg \& Hubermann}{1993}]{Hogg93}
Hogg, T., and Hubermann, B.~A.
\newblock 1993.
\newblock Better than the best: The power of cooperation.
\newblock In Nadel, L., and Stein, D., eds., {\em 1992 Lectures in Complex
  Systems}, volume~V of {\em SFI Studies in the Sciences of Complexity}.
  Reading, MA: Addison-Wesley.
\newblock  163--184.

\bibitem[\protect\citeauthoryear{Kalai}{1990}]{Kalai90}
Kalai, E.
\newblock 1990.
\newblock Bounded rationality and strategic complexity in repeated games.
\newblock In Ichiishi, T.; Neyman, A.; and Tauman, Y., eds., {\em Game Theory
  and Aplications}. San Diego: Academic Press.
\newblock  131--157.

\bibitem[\protect\citeauthoryear{Kaniel}{1994}]{Kaniel94}
Kaniel, R.
\newblock 1994.
\newblock On the equipment rental problem.
\newblock Master's thesis, Hebrew University.

\bibitem[\protect\citeauthoryear{Kreps \bgroup \em et al.\egroup }{1982}]{KMRW}
Kreps, D.; Milgrom, P.; Roberts, J.; and Wilson, R.
\newblock 1982.
\newblock Rational cooperation in finitely repeated {P}risoners' {D}ilemma.
\newblock {\em Journal of Economic Theory} 27(2):245--252.

\bibitem[\protect\citeauthoryear{Megiddo \& Wigderson}{1986}]{Megiddo86}
Megiddo, N., and Wigderson, A.
\newblock 1986.
\newblock On play by means of computing machines.
\newblock In {\em Conference on Theoretical Aspects of Reasoning about
  Knowledge},  259--274.

\bibitem[\protect\citeauthoryear{Mor}{1995}]{MyMast}
Mor, Y.
\newblock 1995.
\newblock Computational approaches to rational choice.
\newblock Master's thesis, Hebrew University.
\newblock In preparation.

\bibitem[\protect\citeauthoryear{Nowak \& Sigmund}{1993}]{NowSig93}
Nowak, M., and Sigmund, K.
\newblock 1993.
\newblock A strategy of win-stay lose-shift that outperforms tit-for-tat in the
  prisoner's dilemma game.
\newblock {\em Nature} 364:56--58.

\bibitem[\protect\citeauthoryear{Papadimitriou}{1992}]{Papadimitriou92}
Papadimitriou, C.~H.
\newblock 1992.
\newblock On players with a bounded number of states.
\newblock {\em Games and Economic Behavior} 4:122--131.

\bibitem[\protect\citeauthoryear{Rapoport \bgroup \em et al.\egroup
  }{1962}]{Rapoport62}
Rapoport, A.; Chammah, A.; Dwyer, J.; and Gyr, J.
\newblock 1962.
\newblock Three-person non-zero-sum nonnegotiable games.
\newblock {\em Behavioral Science} 7:38--58.

\bibitem[\protect\citeauthoryear{Simon}{1969}]{Simon69}
Simon, H.~A.
\newblock 1969.
\newblock {\em The Sciences of the Artificial}.
\newblock Cambridge, Massachusetts: The MIT Press.

\bibitem[\protect\citeauthoryear{Simon}{1983}]{Simon83}
Simon, H.~A.
\newblock 1983.
\newblock {\em Models of Bounded Rationality}.
\newblock Cambridge, Massachusetts: The MIT Press.

\bibitem[\protect\citeauthoryear{Sleator \& Tarjan}{1984}]{Sleator84}
Sleator, D.~D., and Tarjan, R.~E.
\newblock 1984.
\newblock Amortized efficiency of list rules.
\newblock {\em STOC} 16:488--492.

\bibitem[\protect\citeauthoryear{Vanberg \& Congleton}{1992}]{Vanberg92}
Vanberg, V.~J., and Congleton, R.~D.
\newblock 1992.
\newblock Rationality, morality, and exit.
\newblock {\em American Political Science Review} 86(2):418--431.

\end{thebibliography}
\end{document}